\documentclass{elsart}

 \usepackage{epsfig}

\usepackage{amssymb}
\begin{document}
\begin{frontmatter}


\title{ Fingerprints of nonextensive
 thermodynamics in a long-range Hamiltonian system }


\author[label1]
{Vito Latora},   \corauth[cor1]{Corresponding author: vito.latora@ct.infn.it} 
\author[label1]
{Andrea Rapisarda}, 
\author[label2]
{Constantino Tsallis} 

\address[label1] {Dipartimento di Fisica e Astronomia,  Universit\'a di Catania,\\
and INFN sezione di Catania,Corso Italia 57 I 95129 Catania, Italy}

\address[label2] { Centro Brasileiro de Pesquisas Fisicas,  Rua Xavier Sigaud 150,\\
 22290-180
Rio de Janeiro, Brazil}

\begin{abstract}


We study the dynamics of a Hamiltonian system of N classical 
spins with infinite-range interaction. We present numerical results
which confirm   the existence of metaequilibrium Quasi Stationary States (QSS), 
characterized by  non-Gaussian velocity distributions, 
anomalous diffusion, L\'evy walks and dynamical correlation in phase-space.
We show that the Thermodynamic Limit (TL) and the Infinite-Time Limit (ITL) 
do not commute. Moreover, if  the TL  is taken before the 
ITL the system does not relax to the 
Boltzmann-Gibbs equilibrium, but  remains in this {\em new equilibrium state}
where nonextensive thermodynamics seems to apply. 

\end{abstract}

\end{frontmatter}

\section{Introduction}

\label{}

Statistical thermodynamics  is usually intended as the study of 
N-body systems at equilibrium. 
However only a few textbooks \cite{bale} state clearly that 
the validity of equilibrium ensembles as models of thermodynamics 
is not automatically granted, but depends crucially on the nature 
of the Hamiltonian of the N-body system. 
In particular the very same basic postulate of equilibrium statistical 
mechanics, the famous {\em Boltzmann principle} $S=k \log W$ 
 of 
{\it microcanonical ensemble}, assumes that dynamics can
be automatically (and kind 
of easily) taken into account. However this is 
not always justified \cite{einstein,cohen}.
On the other hand,  the Boltzmann-Gibbs {\it canonical ensemble}, 
is valid only for sufficiently short-range interactions 
and does not necessarily apply for example to gravitational or unscreened 
Coulombian fields  for which the  usually assumed 
entropy additivity postulate is not valid \cite{land,fermi}. 
In general, a series of thermodynamic anomalies 
\cite{thir,lyn,pos,gro,dago,ato,kud}, which seem to escape a common 
general framework of understanding, has been observed. 
A few years ago, a generalized thermodynamics formalism based 
on a nonextensive entropy formula 
was proposed \cite{tsa1}. 
The latter has been encountering a large number 
of successful applications in far-from-equilibrium situations, 
as for example, to cite only a few cases among the most recent ones, in 
plasma physics \cite{bog}, heavy-ion collisions \cite{wilk},
turbulence \cite{bec}, discrete maps \cite{maps} 
and even in interdisciplinary fields such as    
bio-physics \cite{idra} and linguistics \cite{monte}.
Such a formalism is the best candidate to 
be the general framework for a thermodynamics
when long-range correlations or fractal structures in phase space 
are important and time evolution is not trivial, 
in other words when {\it the dynamics plays a non trivial
role} \cite{einstein,cohen}. 
In this paper we study the dynamics of relaxation to equilibrium 
in a Hamiltonian system of classical spins with infinite 
range interactions \cite{ant1,lat1,lat2,lat3}.  
We show that, for some values of the initial energy and a class of 
off-equilibrium initial conditions, the systems does not 
relax to the Boltzmann-Gibbs equilibrium, but exhibits 
different equilibrium properties characterized by non-Gaussian 
velocity distributions which can be fitted by   
the probability distribution functions (pdfs) 
of nonextensive thermodyamics\cite{tsa1}. 
The present study, together with the results presented in a 
paper now in press  \cite{lat0}, 
provide the first indication that the generalized 
nonextensive thermodynamics can be a good candidate to 
explain some of the 
anomalies found in hamiltonian systems with long-range 
interactions.

\section{Dynamics and Thermodynamics of the HMF model}

The model, usually called Hamiltonian Mean Field  (HMF), consists  of 
N planar classical spins interacting through an 
infinite-range potential \cite{ant1}. The  Hamiltonian is:
\begin{equation}
        H= K+V
= \sum_{i=1}^N  {{p_i}^2 \over 2} +
  {1\over{2N}} \sum_{i,j=1}^N  [1-cos(\theta_i -\theta_j)]~~,
\end{equation}
\noindent
 where $\theta_i$ is the $ith$ angle and $p_i$ the 
conjugate variable representing the rotational velocity. 
Note that the summation in V is extended to all couples of spins and not
restricted to first neighbors. 
Following tradition, the coupling constant in the potential is 
divided by N. This makes H only formally extensive, i.e. 
$V\propto N$ when $N\rightarrow\infty$\cite{tsa1,celia}, 
since the energy remains non-additive, that is  {\em the system cannot 
be trivially divided in two independent sub-systems}.
The model has an order parameter which is the magnetization
M, i.e.  the modulus of  ${\bf M}={\frac{1}{N}}\sum_{i=1}^N {\bf m}_i$, 
where ${\bf m}_i=[cos(\theta_i), sin(\theta_i)]$.
The canonical analytical solution of the model 
predicts a second-order phase transition from a low-energy 
ferromagnetic phase with magnetization  $M\sim1$,  
to a high-energy one, where the spins are homogeneously oriented 
on   the unit circle and $M\sim0$. The  
dependence of  the energy density $U = E/N$ on the temperature $T$,   usually
called the  {\em caloric curve}, is  given by \cite{ant1,lat1}
\begin{equation}
U = {T \over 2} + {1\over 2} \left( 1 - M^2 \right) ~.
\end{equation}


\begin{figure}
\begin{center}
\epsfig{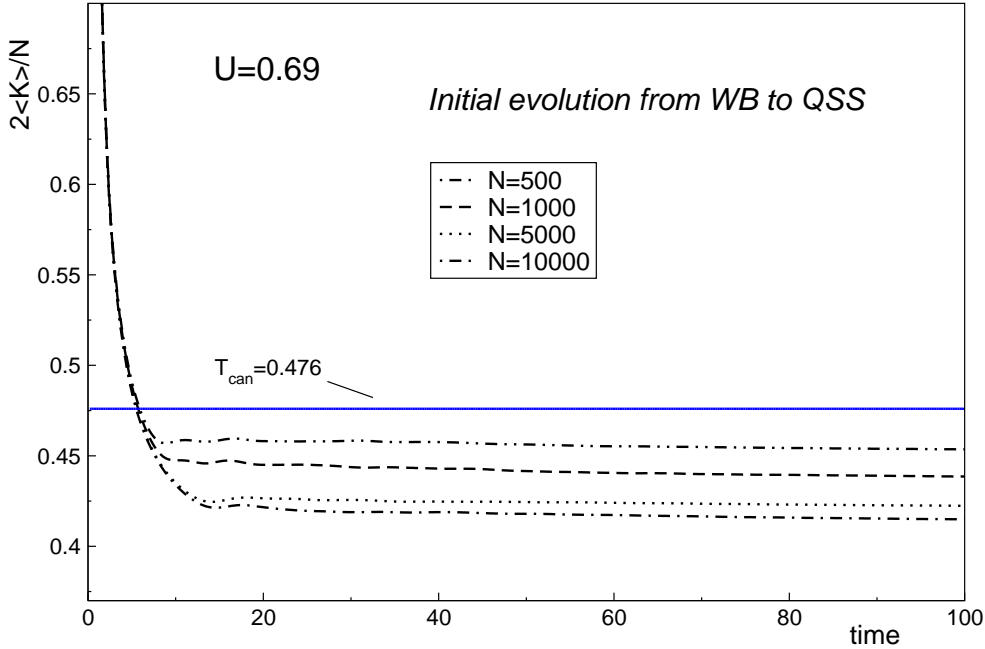}
\end{center}
\caption{  The figure shows the initial time evolution for the Water Bag (WB)
initial state to the QSS for different system sizes. 
The straight full line indicates the
canonical equilibrium temperature. }
\end{figure}

The critical point is at energy density $U_c=0.75$ 
corresponding to a critical temperature $T_c=0.5$ \cite{ant1}. 
The dynamics of HMF can be investigated by starting the 
system with out-of-equilibrium initial conditions 
and integrating numerically the equations of motion \cite{lat1}. 
In particular in refs. \cite{lat1,lat0} we have adopted 
Water Bag (WB) initial conditions, i.e. $\theta_i=0$  for 
all $i$ ($M=1$), and velocities uniformly distributed. 
In a special region of energy values ($0.5<U<U_c$) 
the results of the simulations show, for a transient regime which depends
on the system size, 
a disagreement with the canonical ensemble. In this region  
the dynamics is characterized by  L\'evy walks and 
anomalous diffusion, while  in correspondence 
the system shows a  negative specific heat
\cite{lat3}.
Ensemble inequivalence and negative specific heat 
have also been found in self-gravitating systems \cite{thir,lyn}, 
nuclei and atomic clusters  \cite{gro,dago,ato},
though in the present model such anomalies emerge as dynamical 
features  \cite{ant2,lat4}. 
In this paper we focus on a particular energy value belonging to 
the anomalous region, namely $ U=0.69$, and we study the 
time evolution of temperature, magnetization,  velocity 
distributions.

\begin{figure}
\begin{center}
\epsfig{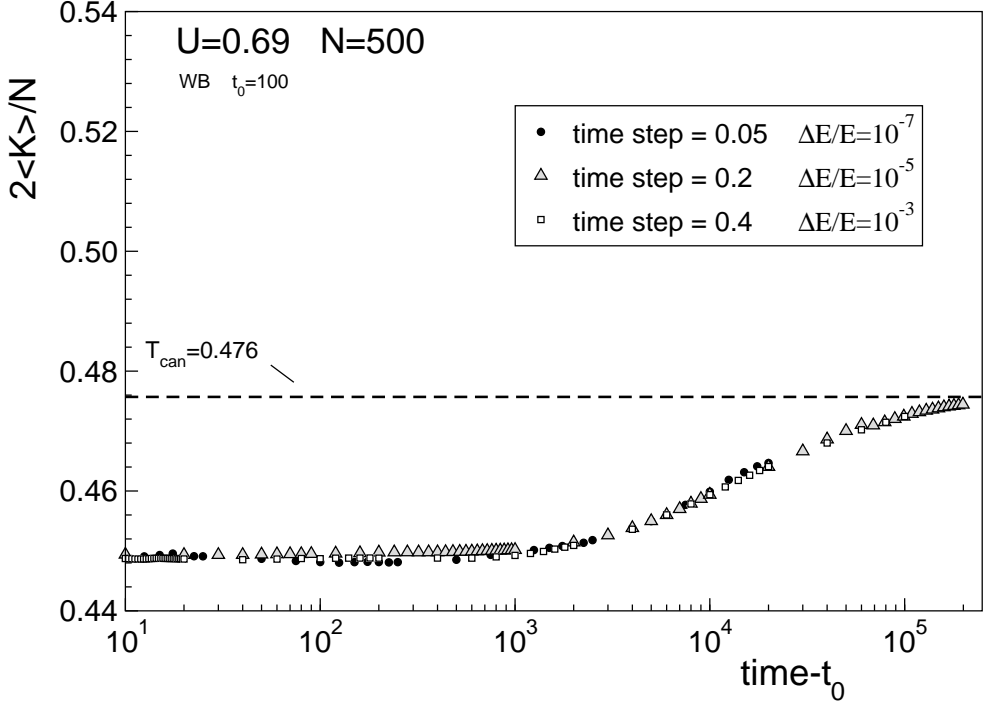}
\end{center}
\caption{ 
For N=500 we show the time evolution of the {\em temperature plateau} 
correspondent to the Quasi Stationary State
(QSS) calculated with different accuracies, reported in the plot. 
The initial time interval $t_0=100$,
shown in Fig.1, has been subtracted for clearness.}
\end{figure}

\begin{figure}
\begin{center}
\epsfig{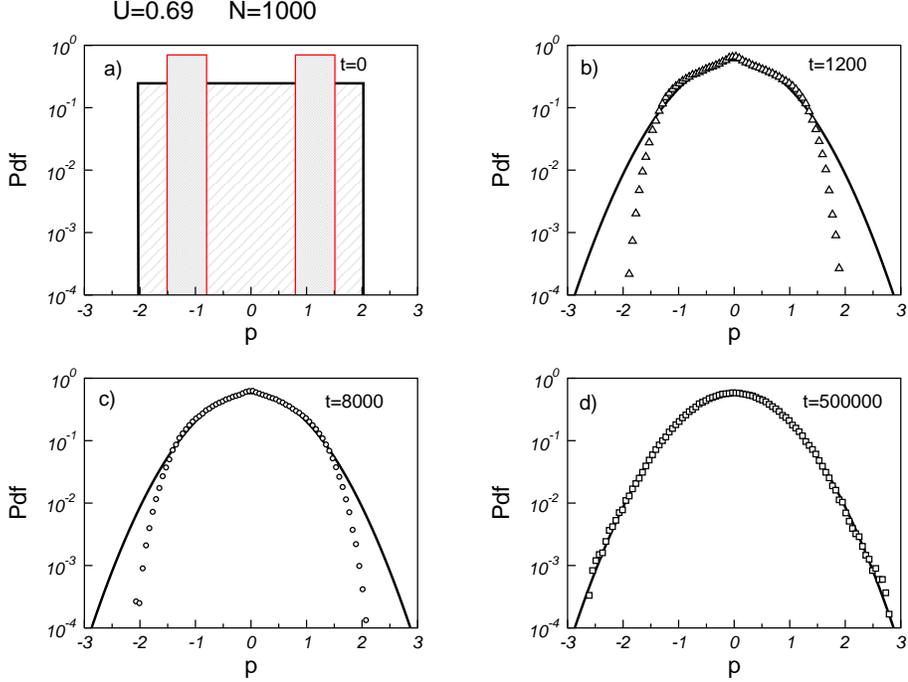}
\end{center}
\caption{ 
Time evolution of numerical velocity pdfs for N=1000 starting 
from WB  and DWB initial conditions, panel a). For comparison we show 
also the equilibrium  Gaussian curve (full line) in the QSS regime,
 panel b-c) (time t=1200,8000),  and in the equilibrium one, panel d) (time t=500000).
}
\end{figure}

In Fig.1  we report the time evolution of $2<K>/N$, a quantity that 
coincides with the temperature 
($<\cdot>$ denotes time averages). 
The system, started with WB initial conditions, 
rapidly reaches a metastable Quasi-Stationary State (QSS) 
which does not coincide with the canonical prediction. 
In fact, after a short transient time, $2<K>/N$ assumes a 
fixed value (the plateau in figure) corresponding to a 
N-dependent temperature $T_{QSS}(N)$  
lower than the canonical prediction, also reported.
In correspondence of this plateau one gets for the magnetization
a value $M_{QSS}\sim 0$.
If we want to observe relaxation to the canonical equilibrium 
state with temperature $T_{can}=0.476$ and magnetization 
$M_{can}=0.307$, we have to wait for a time longer 
than that shown in fig.1, as shown   for example in fig.2 for the case 
$N=500$. 
In ref. \cite{lat0} the following scaling 
relations have been found:
 
\noindent
{\em(i)}  the duration of the plateau, the lifetime of the QSS 
$\tau$,  increases as $\tau \propto N$; 

\noindent
{\em(ii)} $T_{QSS}(N) \rightarrow T_{\infty} =0.380$, a value
obtained analytically as the metastable prolongation,
at energies below $U_c$,  of the high-energy solution ($M=0$)
as $[T_{QSS}(N)-T_{\infty}] \propto N^{-1/3}$. This implies that 
 $M_{QSS} \propto  N^{-1/6}$ (see fig.1(c) of ref. \cite{lat0}).

These numerical results clearly  indicate that 
\begin{itemize}
\item {\em the two limits, the Infinite-Time Limit (ITL)
$ t\rightarrow \infty$ and  the Thermodynamic Limit (TL) $ N\rightarrow \infty$ 
do not commute};  

\item {\em if the TL
is performed before the ITL, 
the system does not relax to the BG equilibrium} and lives forever 
in the QSS. 
\end{itemize}

The robustness 
of the above results was checked in two different ways: 
\noindent
1) by adopting  different initial conditions,  as for example 
double water bag (DWB) initial conditions, and checking that we get
the same QSS (see ref. \cite{lat0} for details);
\noindent
2) by changing the level of accuracy of the numerical integration, 
as shown in fig.2. 
We expect these  results to be  ubiquitous in  
nonextensive systems as conjectured in ref.\cite{tsa1}.

In Fig.3 we study the  velocity pdfs. 
The initial velocity pdfs (WB or DWB initial conditions) 
 quickly acquire and maintain 
during the entire duration of the metastable state 
a  {\em non-Gaussian  shape}. In 
Figs.3(b) and (c) we see  that the pdfs of the QSS do not 
change up to a time $t=8000$ for a system with $N=1000$ .
The velocity pdf of the QSS is wider than a Gaussian 
for small velocities, but shows a faster decrease for $p>1.2$.
The enhancement for velocities around $ p\sim 1$  is   
consistent with the anomalous diffusion and 
the L\'evy walks  
 observed in the QSS regime \cite{lat2}. 
The following rapid decrease for $p>1.2$ is due to conservation of 
total energy. 
From a dynamical point of view, the stability of the QSS velocity pdf 
can be explained by the fact 
that, for $N\rightarrow \infty$, 
  $M_{QSS} \rightarrow 0$  and thus  the force on  
the spins $ F_i  = (-M_x sin\theta_i + M_y cos\theta_i)   \rightarrow 0$. 
On the other hand, when   N is  finite,  we have always 
 a small random force,whose strength depends on N,  
which makes the system  eventually 
evolve into the usual Maxwell-Boltzmann 
 distribution after some time. We show this for N=1000 
at time  t=500000 in fig.3(d).  When this happens, 
  L\'evy walks disappear and anomalous 
diffusion leaves place to  Brownian diffusion \cite{lat2}.

%
%
%
%
\begin{figure}
\begin{center}
\epsfig{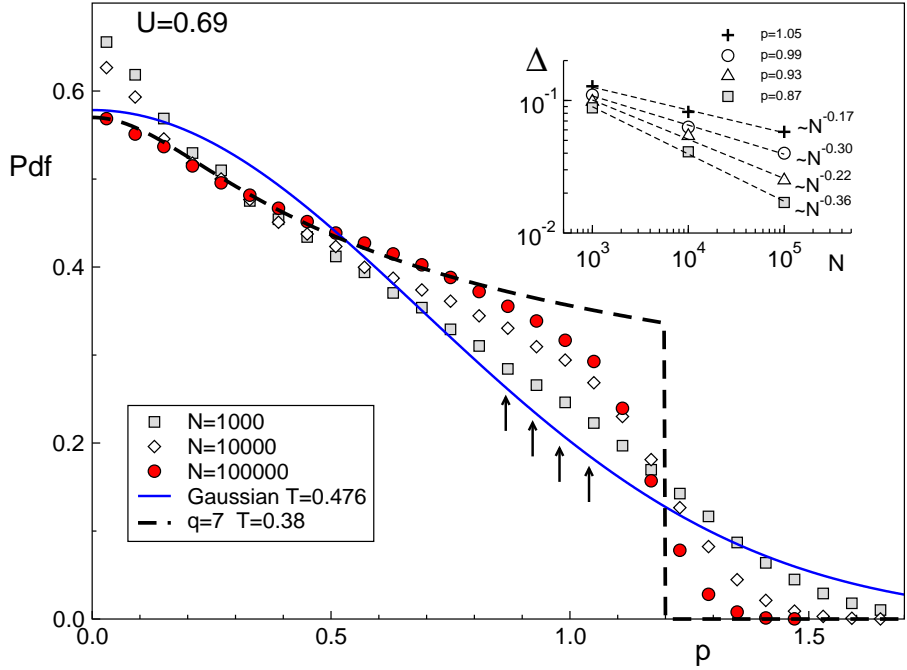}
\end{center}
\caption{ 
Nonextensive theoretical pdf, with cut-off, 
(dashed curve) in comparison with the numerical 
ones (points) and the gaussian equilibrium prediction (full curve). In the inset 
we report the  scaling of $\Delta=P_{th}-P_{num}$ with $N$ for those 
points indicated by the arrows. See text for further details. }
\end{figure}
%
\begin{figure}
\begin{center}
\epsfig{figure=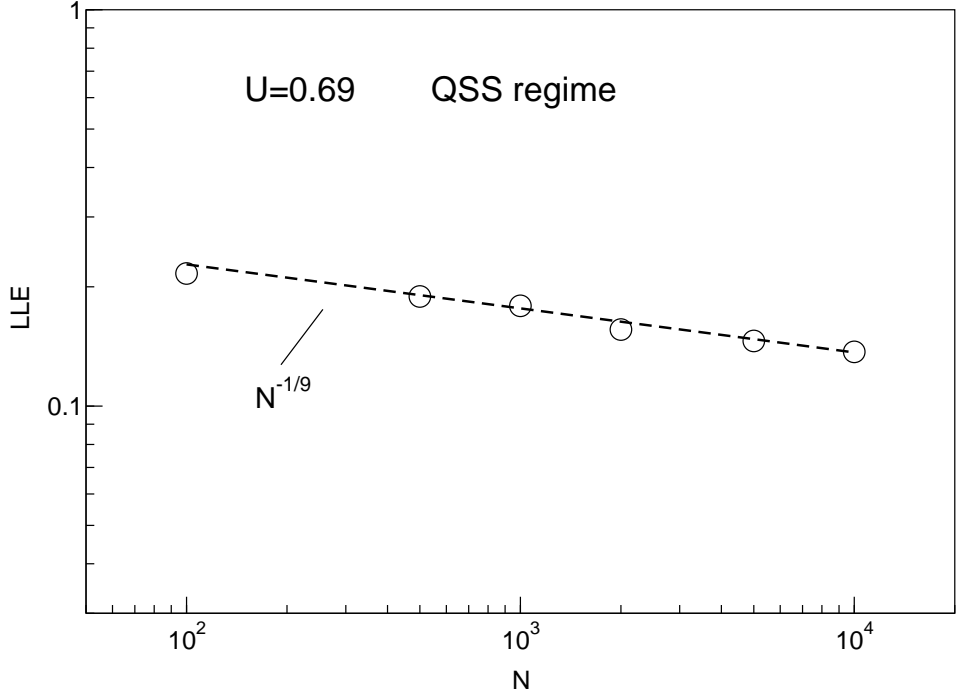,width=11truecm,angle=-90}
\end{center}
\caption{ 
Open points indicate the Largest Lyapunov Exponent (LLE) vs $N$ 
for the QSS regime. An average  over 10 runs is considered. 
Straight line indicates the  theoretical prediction for LLE scaling, see text.
 }
\end{figure}
%
%
%
We fit the non-Gaussian pdf  in fig.3(b) by using the 
one-particle prescription of the generalized thermodynamics
(see ref. \cite{lat0} for more details): 
\begin{equation}  
  P(p) = \left[1 - \left(1 -q \right) {p^2\over2T} \right]^{ 1/(1-q)}  .
\end{equation} 
This formula recovers the Maxwell-Boltzmann distribution for $q=1$
and has been recently used to describe successfully  
turbulence \cite{bec} and non-Gaussian pdfs related 
to anomalous diffusion of  {\em Hydra} cells in
cellular aggregates \cite{idra}. In our case,
the best fit is obtained by a curve with $q=7$, $T=0.38$ 
as shown in Fig. 4. 
The agreement between numerical results and theoretical curve improves 
with the size of the system. A finite-size scaling confirming  
the validity of the fit is reported in 
the inset, where $\Delta=P_{th} - P_{num} $, 
the difference between the theoretical points and the numerical ones,  
is shown to go to zero as a power of N (for four values of $p$). 
Since $q>3$,  the theoretical curve does 
not have a finite integral and therefore it needs to be truncated 
with a sharp cut-off (herein assumed to be discontinuous for simplicity) 
to make the total probability equal to one.
It is however clear that, the fitting 
value $q=7$ is only an effective nonextensive entropic index. 
Although similar non-Gaussian pdfs have  been found previously, it was 
for   dissipative systems 
\cite{bec}, while  this is the first evidence 
in a  Hamiltonian system.

In order to investigate deeper the dynamics of the plateaux observed in figs.1 and 2, we have 
studied the Lyapunov exponent in the QSS regime. 
In fig. 5 we show that, as expected, 
the Largest Lyapunov Exponent (LLE)  
tends to zero when N increases. The scaling behaviour of the $LLE$ 
can be understood following the same argument  
already applied in ref.\cite{lat1} in the overcritical 
region.  It is known in fact that, when the Lyapunov can be estimated by 
the product of random matrices, the  
LLE scales with the power 2/3 of the perturbation \cite{parisi,lat1}. 
In our  case the perturbation is given by the 
magnetization, for which we have the scaling law  $M^2\propto N^{-1/3}$ 
in the QSS regime \cite{lat0}, thus
we get the scaling  $LLE \propto M^{2/3} \propto (N^{-1/3})^{1/3}=N^{-1/9}$.
The latter  is in perfect agreement with the numerical results as shown in the figure.
Since LLE tends to zero  as the system is increasingly large, one can safely say that 
mixing is increasingly slower and the observed anomalies 
in the relaxation process are naturally expected in the sense predicted by Krilov \cite{kry}.
We note finally that,  emergence of dynamical correlations and filamentary sticky structures
in the $\mu$-space have also been observed in  the QSS regime \cite{lat0}.

\section{Conclusions}

In this paper we have studied a simple Hamiltonian system 
with long-range interaction. The dynamics of relaxation process 
is extremely rich and the system shows the 
existence of QSS different from the canonical equilibrium. 
These states  satisfy  the usual attributes of thermal equilibrium 
though  they   systematically differ from what BG statistical mechanics
has made familiar to us. 
Our results provide a first verification of 
nonextensive statistical mechanics \cite{tsa1} 
in long-range Hamiltonian systems,  
and illustrate the correctness of the criticism that
Einstein developed in his celebrated 1910 paper \cite{einstein} 
about the possible
nonuniversality of Boltzmann thermostatistics, and the need of providing a
mechanical basis to it (including naturally {\it time} in the discussion).

\vskip 0.25truecm
\noindent
We thank S. Ruffo for the suggestion of the LLE scaling in 
the QSS regime, and E.G.D. Cohen for stimulating discussions.

\bigskip

\noindent

\end{document}